\documentclass[prb,preprint]{revtex4} 

\usepackage{amsmath}  
\usepackage{amsfonts} 
\usepackage{graphicx} 

\usepackage{dsfont} 

\newcommand{\omt}{\tilde\omega}
\newcommand{\inv}{\mathcal I}
\newcommand{\hilb}{\mathcal H}
\newcommand{\rbb}{\mathbb R}

\newcommand{\ra}{{\mathbf r}_A}
\newcommand{\rb}{{\mathbf r}_B}
\newcommand{\rvec}{{\mathbf r}}
\newcommand{\Rvec}{{\mathbf R}}

\newcommand{\be}{\begin{equation}}
\newcommand{\ee}{\end{equation}}

\begin{document}

\title{The van der Waals interaction in one, two, and three dimensions}

\author{A. C.~Ipsen}
\affiliation{Niels Bohr Institute, University of Copenhagen, Blegdamsvej 17, 2100 Copenhagen {\O}, Denmark}

\author{K.~Splittorff}
\affiliation{Discovery Centre, Niels Bohr Institute, University of Copenhagen, Blegdamsvej 17, 2100 Copenhagen {\O}, Denmark}

\date{\today}

\begin{abstract}
The van der Waals interaction between two polarizable atoms is 
considered. In three dimensions the standard form with 
an attractive $1/|\Rvec|^6$ potential is obtained from second-order quantum
perturbation theory. When the electron motion is restricted to lower
dimensions (but the $1/|\Rvec|$ Coulomb potential is retained), new terms in the expansion 
appear and alter both the sign and the $|\Rvec|$ dependence of the interaction.   
\end{abstract}

\maketitle 

\section{Introduction}

One of the most beautiful applications of perturbation theory
in quantum mechanics is the computation of the van der Waals
force between two atoms. The very existence of this force is
perhaps surprising, because it arises even when both atoms are electrically neutral,
resulting from the fact that the two atoms
polarize each other. The simplest example is the interaction between 
two hydrogen atoms. If the two atoms are far apart the 
interaction between them is negligible and in the ground state the 
electron wave functions are spherically symmetric. As the two atoms approach each other
a correlation between the atomic states arises, causing the van der Waals force.
At leading order in the inverse
distance between the atoms this is a dipole-dipole interaction. 
For a general introduction see Refs.~\onlinecite{TMF,Milton,SKSF}. 

\begin{figure}[t*]
\includegraphics[width=8.5cm,angle=0]{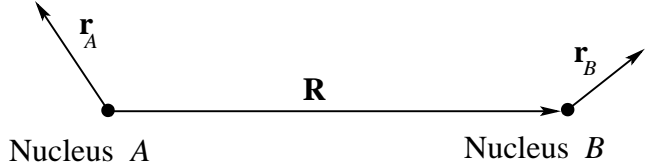}
\caption{\label{fig:R} The nuclei of two hydrogen atoms, indicated by the dots, are 
placed a distance $|\Rvec|$ apart. The 
associated positions of the electrons are at the end of the vectors $\ra$ and $\rb$. While each atom is electrically neutral their polarizability gives rise to the van der Waals interaction. }
\end{figure}

The computation 
of the  strength of the van der Walls
interaction is a classic problem in quantum mechanics.
The problem was originally treated in Refs.~\onlinecite{EL,London,London2},
and can be formulated as follows:

\begin{quote}
Two hydrogen atoms are 
separated by a distance $|\Rvec|$; see Fig.~\ref{fig:R}. 
Use perturbation theory to compute the correction to the ground 
state energy of the system due to the atoms' polarizability, for large 
values of $|\Rvec|$.
\end{quote}

\noindent
The problem is particularly appealing from the pedagogical 
perspective. It challenges the student to understand the 
central concepts of perturbation theory in a physically 
relevant case: What is the unperturbed system?
What is perturbing Hamiltonian? What 
order in the perturbation is needed, and how large must $|\Rvec|$ be 
in order for the results to apply?

Versions of the above problem are standard in quantum mechanics 
references.\cite{LL,Schiff,Sakurai,Kittel,BJ,Abers,Griffiths,Holstein,Stone,AM,Patterson,Born} 
In order to make the problem less technical for the students, 
it is tempting to simplify the computations by constraining the 
motion of the electrons to one or two dimensions, 
\cite{Kittel,Griffiths,Holstein,Stone,Patterson,Born} while keeping the 
$1/|\Rvec|$ Coulomb interaction between the ``atoms'' (we will continue to use the
word ``atom'' even when electrons are constrained to lower dimensions). 
Here we point out that this
reduction also offers a good opportunity to discuss one of the common 
pitfalls of perturbation theory, namely the potential inconsistency of the 
perturbation series; this pitfall appears to have been overlooked in 
Refs.\ \onlinecite{Born,Kittel,Patterson,Griffiths,Holstein,Stone}.

The consistency of the perturbation series comes into play because
the problem contains two expansions: that of the interaction Hamiltonian
and that of (quantum mechanical) perturbation theory.
The interaction Hamiltonian is given by the Coulomb interaction between
the two atoms (each consisting of an electron and a nucleus). 
This interaction vanishes rapidly for large $|\Rvec|$ because both atoms are
neutral. It is therefore natural to expand the interaction Hamiltonian
in inverse powers of $|\Rvec|$ (this can be understood as a multipole expansion).
The leading-order term in the expansion of the interaction Hamiltonian
is of order $1/|\Rvec|^3$, and
the familiar van der Waals term of order $1/|\Rvec|^6$ arises in
perturbation theory from the second-order contribution of this
$1/|\Rvec|^3$ term. Because the first-order contribution from the $1/|\Rvec|^3$
term is zero, it is tempting to conclude that the familiar van der
Waals term of order $1/|\Rvec|^6$ is the leading term in the perturbative
expansion.

However, as we will show in detail below, the $1/|\Rvec|^5$ term of the
perturbing Hamiltonian has a non-vanishing ground-state expectation value
in one and two dimensions. This results in a leading term of order
$1/|\Rvec|^5$. We stress again that this is still for a $1/|\Rvec|$ interaction. Only in three dimensions, as explicitly
noted by Refs.\ \onlinecite{EL,LL,Schiff,Sakurai}, does the ground-state expectation value of the
$1/|\Rvec|^5$ term of the perturbing Hamiltonian vanish, so the leading
term is the second-order correction due to the $1/|\Rvec|^3$ term of the
perturbing Hamiltonian. The standard 
$1/|\Rvec|^6$ dependence of the van
der Waals interaction is therefore special to three dimensions.

\section{The system}

We consider the nuclei of the atoms fixed, so the Hilbert space is just the
product of the Hilbert space of each electron:
\be
  \hilb = \hilb_A\otimes\hilb_B = L^2(\rbb^d)\otimes L^2(\rbb^d),
\ee
where $d\geq 1$ is the number of space dimensions that the electrons are allowed to explore.
The total Hamiltonian is
\be
  H = H_0+H_I,\qquad H_0 = H_A + H_B,
\ee
where $H_A$ and $H_B$ describe the two atoms and the interaction is\cite{Note1}
\be
  H_I = k\left(\frac{1}{|\Rvec|}+\frac{1}{|\Rvec-\ra+\rb|}-\frac{1}{|\Rvec-\ra|}-\frac{1}{|\Rvec+\rb|}\right),
  \label{eq:HI}
\ee
with
\be
  k = \frac{e^2}{4\pi\epsilon_0}.
  \label{eq:k-def}
\ee
(See Fig.~\ref{fig:R}.)
In three dimensions this is the standard form of the Coulomb interaction. 
Let us emphasize that we do not assume that the atom Hamiltonians $H_A$ and $H_B$ are
of the usual hydrogen form (see also the discussion around Eq.~\eqref{eq:hydrogen}).

Below we consider 
also the situation where the motion of the electrons is confined to one or two dimensions, 
but we keep the form of the interaction Hamiltonian. In other words the electromagnetic 
field is always allowed to explore all three spatial dimensions. 

Note that we ignore the fermionic nature of the electrons. The model therefore makes sense
only when $|\Rvec|$ is much larger than the size of the atoms, so the overlap of the electron 
wavefunctions is negligible. Instead of hydrogen atoms, we could more generally consider 
single-valence-electron atoms, i.e., (neutral) atoms with a single electron outside a closed 
shell.

Since the general form of the van der Waals interaction in one, two, and three dimensions 
follows from the symmetries, we do not need the explicit form of $H_A$ and $H_B$.
Rather we simply assume that $H_A$ and $H_B$ have rotational symmetry.
In more detail, let $U_{M,A}$ be the unitary rotation operator defined by
\be
  (U_{M,A}\psi)(\ra,\rb) = \psi(M\ra,\rb),
\ee
where $M$ is an $SO(d)$ matrix for $d\geq 2$, and for $d=1$ we set $M=\pm 1$.
By rotational symmetry we mean that $U_{M,A}$ (and the corresponding $U_{M,B}$) 
commutes with $H_0$. We further assume that the ground state $\psi_0$ (we will 
also use the notation $|0\rangle$) of $H_0$ is unique (and that degeneracies due to other
degrees of freedom, like spin, are irrelevant). It follows that
\be
  |\psi_0(M\ra,M'\rb)| = |\psi_0(\ra,\rb)|.
\ee
For notational simplicity we assume that the two atoms are identical.

Before we proceed with the calculation, let us note that for sufficiently large $|\Rvec|$ 
the form  of the interaction \eqref{eq:HI} is modified by QED effects.\cite{CP}
We will not discuss this complication further, but refer the interested reader to, for example,
Ref.~\onlinecite{Holstein}.

\section{General form of the van der Waals interaction}

To derive the form of the van der Waals interaction we will assume that 
$|\Rvec|$ is sufficiently large and compute the correction to the ground state 
energy using perturbation theory. 
As mentioned in the introduction this is a standard problem in quantum mechanics,
see e.g.~Refs.~\onlinecite{LL,Schiff,Sakurai,Kittel,BJ,Abers,Griffiths,Holstein,Stone,AM,Born,Patterson}, 
however when constraining the motion of the electrons to one or two dimensions the nature
of the interaction changes.
First we will show how the familiar 
$1/|\Rvec|^6$ term arises and subsequently, by carefully checking the consistency 
of the perturbative expansion, we will show that in one and two dimensions 
the $1/|\Rvec|^6$ term is subleading.

\subsection{The familiar $1/|\Rvec|^6$ form}

To calculate the correction to the ground state energy of the full system, we 
expand $H_I$ in powers of $1/|\Rvec|$. The leading term is of order $1/|\Rvec|^3$
\be
\label{HIexp}
  H_I = k \frac{(\ra\cdot \rb)|\Rvec|^2-3(\Rvec \cdot \ra)(\Rvec \cdot \rb)}{|\Rvec|^5}+O(|\Rvec|^{-4}).
\ee
Let us choose coordinates such that $\Rvec$ points along the $x$-axis. Since 
we have assumed that the ground state is unique we may simply plug this into the 
standard formula for the first order correction. We then find ($E_n$ denotes
the energy levels of the \emph{unperturbed} Hamiltonian $H_0$)
\be
  \Delta_1 E_0 = \frac{k}{|\Rvec|^3}\langle 0 |\ra \cdot \rb-3x_Ax_B|0\rangle + O(|\Rvec|^{-4}).
  \label{eq:first-order}
\ee
Here $|0\rangle$ is the ground state of $H_0 = H_A+H_B$ which has the product form
\be
  |0\rangle = |0_A\rangle|0_B\rangle,
\label{productform}
\ee
and by rotational symmetry, we have
\be
  \langle 0_A|\ra|0_A \rangle = \langle 0_B|\rb|0_B \rangle = \mathbf{0}.
\ee
Thus 
\be
  \Delta_1 E_0 = 0 + O(|\Rvec|^{-4}).
\ee

Let us go on to the second order correction due to the $1/|\Rvec|^3$ term in $H_I$:
\begin{eqnarray}
\label{Delta2}
  \Delta_2 E_0 &=& -\sum_{n\neq 0}\frac{|\langle n |H_I|0\rangle|^2}{E_n-E_0}\label{eq:D2}\\
  &=& -\frac{k^2}{|\Rvec|^6}\sum_{n\neq 0}\frac{|\langle n |\ra \cdot \rb-3x_Ax_B|0\rangle|^2}{E_n-E_0}
    + O(|\Rvec|^{-8}), \nonumber
\end{eqnarray}
where $|n\rangle$ are the eigenstates of $H_0$. As we have indicated, there is no 
$1/|\Rvec|^7$ term, see Appendix \ref{app:no-seven} for details. Now one might be satisfied, since we 
have reproduced the expected $1/|\Rvec|^6$ attractive potential (note that the sum is positive),
but as we shall now see the $1/|\Rvec|^6$ is only the leading term in 3 spatial dimensions.

\subsection{Consistency of the expansion and the difference between one, two and three dimensions}
\label{sec:consistency}

The second order correction, $\Delta_2 E_0$, to the ground state energy we found 
is $O(|\Rvec|^{-6})$. We used second order perturbation theory since the first order term
vanished $\Delta_1 E_0 = 0$. However, for the first order correction we used an
expansion of the interaction 
Hamiltonian, Eq.~(\ref{HIexp}), which only holds up to $O(|\Rvec|^{-4})$. So 
in order to check the consistency of the expansion we should 
calculate the first order corrections also due to terms up to order $1/|\Rvec|^6$ in $H_I$. 

Instead of simply expanding $H_I$ and calculating the expectation value, we will follow
a slightly indirect  route, which will prove more enlightening
(the completely equivalent standard approach is included in Appendix \ref{app:direct}). 
To this end, first note that
\be
  \Delta_1E_0 = \langle 0|H_I|0\rangle = k\langle 0_B|V_A(\Rvec)-V_A(\Rvec+\rb)|0_B\rangle,
  \label{eq:D1-VA}
\ee
with
\be
  V_A(\rvec) = \left\langle 0_A\middle| \frac{1}{|\rvec|}-\frac{1}{|\rvec-\ra|} \middle|0_A \right\rangle.
\ee
Writing $V_A(\rvec)$ as 
\be
  V_A(\rvec) =  \int d^d\ra \frac{\delta^d(\ra)-|\psi_{0,A}(\ra)|^2}{|\rvec-\ra|},
  \label{eq:VA-int}
\ee
it is clear that it is (proportional to) the electrostatic potential of the ground state
of the $A$ atom.

We expand (this is just the familiar multipole expansion),
\be
  V_A(\rvec) = \left\langle 0_A\middle| -\frac{\rvec \cdot \ra}{|\rvec|^3}+\frac{|\ra|^2}{2|\rvec|^3}
             -\frac{3(\rvec \cdot \ra)^2}{2|\rvec|^5}+O(|\rvec|^{-4})\middle|0_A \right\rangle
  \label{eq:VA-exp}
\ee
and using
\be
 \langle 0_A| |\ra|^2 |0_A \rangle = \frac{d\langle 0_A| (\rvec \cdot \ra)^2 |0_A \rangle}{|\rvec|^2}
\ee
we find that in general (note that $V_A$ only depends on $|\rvec|$ by rotation symmetry)
\be
  V_A(\rvec) = -\frac{(3-d)a^2}{2}\frac{1}{|\rvec|^3}+O(|\rvec|^{-5}).
  \label{eq:VA-exp2}
\ee
Here the characteristic length $a$ is defined by
\be
  a^2 = \frac{\langle 0_A| |\ra|^2 |0_A\rangle}{d}.
  \label{eq:a-def}
\ee
Any $|\rvec|^{-4}$ term in \eqref{eq:VA-exp} would have to contain three
factors of $\ra$, and would thus vanish by the inversion symmetry.

We now plug the result for $V_A$ into \eqref{eq:D1-VA}, and find
\be
  \Delta_1E_0 = -\frac{3(3-d)k a^2}{2}\left\langle 0_B\middle|\frac{\Rvec \cdot \rb}{|\Rvec|^5}
  +\frac{|\rb|^2}{2|\Rvec|^5}-\frac{5 (\Rvec \cdot \rb)^2}{2|\Rvec|^7}
  +O(|\Rvec|^{-6})\middle|0_B \right\rangle,
\ee
so the general first order correction takes the form (even negative powers again vanish by
inversion symmetry)
\be
  \Delta_1E_0 = \frac{3(3-d)(5-d)k}{4}\frac{a^4}{|\Rvec|^5} +O(|\Rvec|^{-7}).
\label{Delta1E0}
\ee
For the standard derivation of this result see Appendix \ref{app:direct}. 
We conclude that for $d = 1,2$  a \emph{repulsive} $1/|\Rvec|^5$ term is present, which we had
missed before. However, for $d=3$ the leading term is indeed the $1/|\Rvec|^6$ term. This
demonstrates the importance of the consistency of the expansion.\cite{Note2}

The $1/|\Rvec|^5$ will also become the dominant term in 3 spatial dimensions provided that we 
consider corrections to \emph{excited} states of the atoms (degeneracies can even change 
it to $1/|\Rvec|^3$), see e.g.~Refs.~\onlinecite{LL} and \onlinecite{Abers}.

In the preceding discussion we have tacitly assumed that $a^2 > 0$ such
that the $1/|\rvec|^3$ term in $V_A(\rvec)$ is non-zero, cf. Eq. \eqref{eq:VA-exp2}.
Is it possible to come up with (singular) models that violate this? If we let
the atom Hamiltonian take the hydrogen like form
\be
  H_A = -\frac{\hbar}{2m}\frac{d^2}{d x_A^2} - \frac{k}{|x_A|}
  \label{eq:hydrogen}
\ee
in $d=1$ the ground state wavefunction becomes completely localized at $x_A = 0$, see
Ref.~\onlinecite{Loudon}.
We thus have $a^2 = 0$ and hence no $1/|\Rvec|^5$ correction to $E_0$. 
In the following we will focus on the generic $a^2 > 0$ situation.

\section{Geometrical interpretation}

We have seen that that $V_A(\rvec)= 0 + O(|\rvec|^{-5})$ for $d=3$. There is a simple way to understand this.
Consider a rotationally symmetric
distribution $\rho$ in $d=3$ with bounded support, in the sense that 
$\rho(\rvec)=0$ for $|\rvec|> r_*$. From electrostatics we know that
\be
  \int d^3 \rvec' \frac{\rho(|\rvec'|)}{|\rvec-\rvec'|} = \frac{q}{|\rvec|},\qquad
   q=\int d^3 \rvec' \rho(|\rvec'|) 
\ee
when $|\rvec| > r_*$. Looking back at \eqref{eq:VA-int}, it follows immediately that (in $d=3$)
\be
  V_A(\rvec) = \frac{1}{|\rvec|}-\frac{1}{|\rvec|} = 0
\ee
as long as $\rvec$ is outside the atom. Under the assumption that electron wavefunctions of the two
atoms don't overlap (which is necessary for the consistency of the model anyway), we conclude 
that the first order correction 
$\Delta_1E_0$ vanishes to all orders in $|\Rvec|^{-1}$ in $d=3$. This can also be understood in
terms of the orthogonality properties of the spherical harmonics, see 
e.g.~Refs.~\onlinecite{Schiff} and \onlinecite{Sakurai}.
(Of course, for physically realistic
wavefunctions there will be some
overlap, but it will fall off at least as fast as $\exp(-|\Rvec|/a)$, 
which will not show up in an expansion in $1/|\Rvec|$, see Ref.~\onlinecite{LL}.)

\begin{figure}[t*]
\includegraphics[width=8.5cm,angle=0]{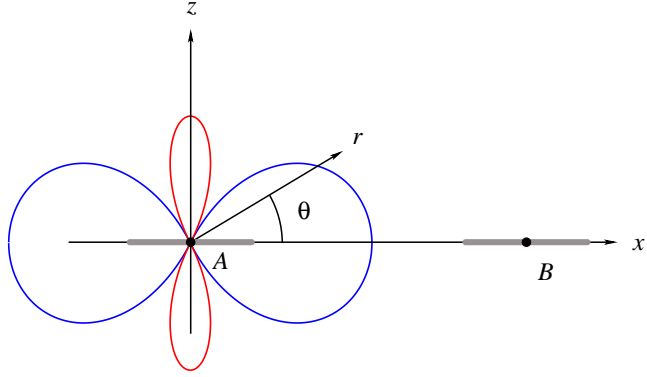}
\caption{\label{fig:geom} The one dimensional system embedded in three dimensional space.  Here we only show the $xz$ plane. The one dimensional electron `clouds' are indicated by the horizontal gray lines. A polar plot of \eqref{eq:quadro} shows the shape of the $A$ atom's quadrupole. On the two lobes along the $z$-axis $V_A(\rvec)$ is positive, while
it is negative on the lobes along the $x$-axis.  
For the two dimensional system the picture would be similar, but with the electron clouds forming discs in the $xy$ plane.}
\end{figure}

To understand why the atoms repel each other for $d<3$ it is useful to think of the
system as embedded in three dimensional space. It is then meaningful to ask what
the  potential $V_A(\rvec)$ is,  if $\rvec$ is allowed to be a three dimensional vector. 
Let us set
\be
  \rvec = |\rvec|(\hat {\mathbf x} \cos\theta + \hat {\mathbf z} \sin\theta),
  \label{eq:three-d-r}
\ee
where $\hat {\mathbf x}$ is a unit vector parallel to the system, while $\hat {\mathbf z}$
is perpendicular to the
system, see Figure \ref{fig:geom}. If we plug \eqref{eq:three-d-r} into \eqref{eq:VA-exp}
we find
\be
  V_A(\rvec) = -\frac{(3\cos^2\theta-d)a^2}{2}\frac{1}{|\rvec|^3}+O(|\rvec|^{-5}),\qquad \text{for $d=1,2$},
  \label{eq:quadro}
\ee
which reduces to \eqref{eq:VA-exp2} when $\theta = 0$ (or $\theta = \pi$) as it should.
We recognize the leading term as the potential of a quadrupole (the 
symmetry of the problem excludes the appearance of dipoles). In the model the quadrupoles
of the atoms are aligned such that they will repel each other. The repulsion can 
thus be understood as the result of the permanent quadrupole moments of the atoms.

\section{The Drude model}

Above we have provided the general from of the van der Waals interaction 
based on symmetry arguments. Here we exemplify the general results 
in a simple model for the atoms: the Drude model. In the Drude model, see 
e.g.~Refs.~\onlinecite{Kittel,Griffiths,Stone,Born,London2,Patterson,Holstein}, the electrons 
are bound by a harmonic potential,
\be
  H_{A} = -\frac{\hbar\nabla^2_{A}}{2m}+\frac 1 2 m\omega^2|\rvec_{A}|^2,\qquad
  H_{B} = -\frac{\hbar\nabla^2_{B}}{2m}+\frac 1 2 m\omega^2|\rvec_{B}|^2.
\ee
If we only keep the leading term of $H_I$ given in Eq.~(\ref{HIexp}), the Drude 
model remains harmonic, and we can write down the `exact' correction to the 
ground state energy, see
 e.g.~Refs.~\onlinecite{Kittel,Griffiths,Stone,Born,London2,Patterson,Holstein}.
To do this one changes coordinates to $\rvec_{\pm} = (\ra\pm \rb)/\sqrt{2}$. In terms
of $\rvec_{\pm}$ the model (with the truncated $H_I$) is just $2d$ decoupled oscillators, 
and the ground state energy is
\be
  \Delta E_0 = \frac{\hbar}{2}\left(\omt_{2}+\omt_{-2}+(d-1)(\omt_1+\omt_{-1}) -2d\omt_0\right),
  \label{eq:Drude-exact}
\ee
with the shifted frequencies (note that $k=e^2/4\pi\epsilon_0$ is not the spring constant)
\be
  \omt_n^2 = \omega^2 + \frac{nk}{m|\Rvec|^3}.
\ee

If we expand \eqref{eq:Drude-exact} in $1/|\Rvec|$ we obtain (there is no $1/|\Rvec|^3$ or $1/|\Rvec|^9$ term because
\eqref{eq:Drude-exact} is symmetric under $|\Rvec|\to -|\Rvec|$)
\be
  \Delta E_0 = -\frac{(3+d)k^2 a^4}{2\hbar\omega}\frac{1}{|\Rvec|^6}+O(|\Rvec|^{-12}).
  \label{eq:Drude-D2}
\ee
Here $a$, as defined by \eqref{eq:a-def}, is
\be
   a^2 = \frac{\hbar}{2m\omega}.
   \label{eq:a-Drude}
\ee
It is easy to check that we get the same result from \eqref{eq:D2}, i.e.~from the second 
order correction due to the $1/|\Rvec|^3$ term in $H_I$. In 3 dimensions this is thus completely 
self consistent and provides the leading order correction due to the interaction between 
the two neutral atoms. However, in one and two dimensions, as we have seen above, the leading 
term is of order 
$1/|\Rvec|^5$: The general form of the leading corrections to the Drude ground state energy is
\be
  \Delta E_0 = \frac{3(3-d)(5-d)k}{4}\left(\frac{a^4}{|\Rvec|^5}+\frac{5(7-d)a^6}{2|\Rvec|^7}\right)
  -\frac{(3+d)k^2 a^4}{2\hbar\omega}\frac{1}{|\Rvec|^6}+O(|\Rvec|^{-8}).
\ee
Here we have included the $1/|\Rvec|^7$ term, see Appendix \ref{app:first-order-seven} for details.

\begin{figure}[t*]
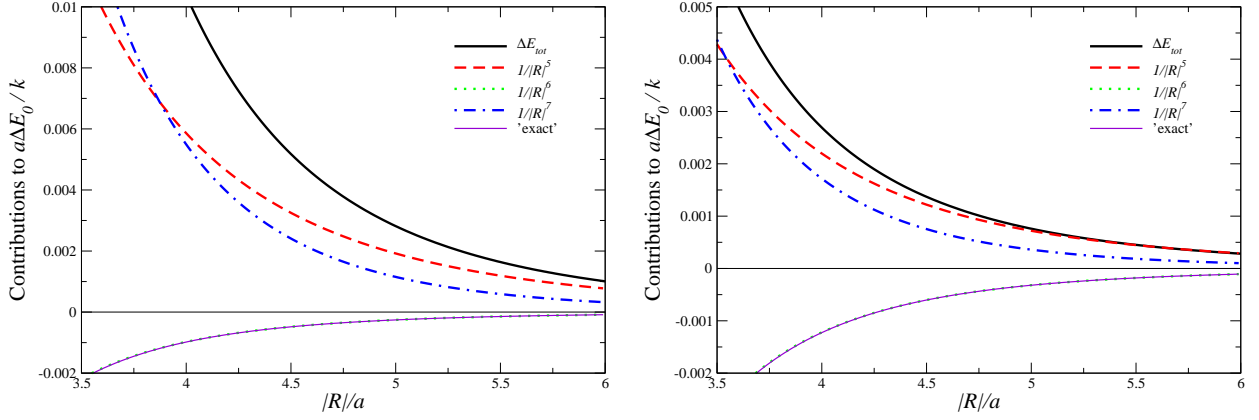

\includegraphics[width=8cm,angle=0]{IpsenSplittorffFig03.eps}
\hfill\includegraphics[width=8cm,angle=0]{IpsenSplittorffFig04.eps}
\caption{\label{fig:DeltaE} The corrections to the ground state energy in the Drude model 
{\bf left:} in 1 dimension  {\bf right:} in 2 dimensions. The thick line shows the total 
 correction up to 
order $1/\tilde{R}^7$. The curve labeled `exact' is a plot of Eq.~\eqref{eq:Drude-exact}. 
 The contributions from the terms of order $1/\tilde{R}^5$ (the dashed lines) are dominant
  for $\tilde{R} \gtrsim 5$.
Note that the `exact' curve is indistinguishable from $1/\tilde{R}^6$ curve in both plots.
This is to be expected since they only start to differ at $O(\tilde{R}^{-12})$, see 
Eq. \eqref{eq:Drude-D2}.
} \end{figure}

To get a feeling for the magnitude of the different terms, let us choose $m$ to be the electron mass,
$m_e$, and $a$ to be the Bohr radius
\be
  a = a_{\text{Bohr}} = \frac{\hbar^2}{m_e k}.
\ee
Using \eqref{eq:a-Drude} we then have
\be
  \frac{a\Delta E_0}{k} = \frac{3(3-d)(5-d)}{4}\left(\frac{1}{\tilde R^5}
  +\frac{5(7-d)}{2}\frac{1}{\tilde R^7}\right)
  -(3+d)\frac{1}{\tilde R^6}+O(\tilde R^{-8}),
\ee
where the dimensionless separation is
\be
  \tilde R = \frac{|\Rvec|}{a}.
\ee
The correction is plotted in Figure \ref{fig:DeltaE} for one and two dimensions. 
Note that we should only trust
our perturbative calculations in the region where the `leading' $1/|\Rvec|^5$ correction dominates
the other corrections, i.e. for $\tilde R \gtrsim 5$.

\section{Conclusion}

The van der Waals interaction between two neutral atoms offers a 
perfect exercise in quantum mechanics. It allows the 
student to gain experience with the basic concepts of perturbation 
theory in a physically relevant case. The problem also naturally 
suggests itself to discuss more advanced concepts such as 
retardation.\cite{Holstein}  Here we have used the van der 
Waals interaction to emphasize the importance of the consistency of 
the perturbation series. In particular, we have shown that 
when the electron motion is restricted to one and 
two dimensions the ordering of of the perturbative series is different 
from the familiar form obtained in three spatial dimensions. This affects 
the $|\Rvec|$-dependence of the van der Waals interaction which becomes a
repulsive and of order $1/|\Rvec|^5$ in one and two dimensions. This pitfall
offers a great chance to discuss the importance of the consistency of 
the perturbation series and appears to have been overlooked in the 
literature.\cite{Kittel,Griffiths,Holstein,Stone} 

The presentation has been based almost entirely on symmetry arguments.  
Explicit evaluation of the perturbation series has been presented for 
the Drude model. We hope that this discussion may serve as inspiration 
also at other universities and colleges.

\begin{acknowledgments}
It is a pleasure to thank colleagues and students at the Niels Bohr 
Institute for useful discussions. We also thank the anonymous
referees for helpful comments and additional references.
The work 
of KS was supported by the {\sl Sapere Aude program} of The Danish 
Council for Independent Research. The work of ACI was supported by
the ERC-Advanced grant 291092 ``Exploring the Quantum Universe''.
\end{acknowledgments}

\appendix  
\section{No $1/|\Rvec|^7$ dependence of $\Delta_2E_0$}
\label{app:no-seven}

Here we explain why a $1/|\Rvec|^7$ dependence is excluded from $\Delta_2E_0$
in Eq.~(\ref{Delta2}). Consider the unitary inversion operator, $\inv$, defined by
\be
  (\inv\psi)(\ra,\rb) = \psi(-\ra,-\rb).
\ee
It is clear that
\be
  \inv \rvec_{A/B} \inv = -\rvec_{A/B}, \qquad \inv^2 = \mathds{1}.
\ee
Now $\inv$ commutes with $H_0$,\cite{Note3}
which means that we can assume that the $|n\rangle$ are also eigenstates of $\inv$. We can
thus split the
sum in \eqref{eq:D2} as
\be
  \Delta_2 E_0 = -\sum_{\substack{n\neq 0\\ \inv|n\rangle = +|n\rangle}}
    \frac{|\langle n |H_I|0\rangle|^2}{E_n-E_0}
  -\sum_{\substack{n\neq 0\\ \inv|n\rangle = -|n\rangle}}
    \frac{|\langle n |H_I|0\rangle|^2}{E_n-E_0}.
\ee
It is  easy to see that the $|\Rvec|^{-3}$ terms of $H_I$ will only contribute to the
first sum, while the $|\Rvec|^{-4}$ terms will only contribute to the second sum. Hence, 
a term of order $1/|\Rvec|^7$ cannot result.

\section{Direct calculation of $\Delta_1 E_0$}
\label{app:direct}

In Section \ref{sec:consistency} we calculate the first order correction in an
indirect way by first considering the potential $V_A(\rvec)$. Here we outline
a more standard brute force derivation of \eqref{Delta1E0}. With 
$\Rvec = |\Rvec|\hat {\mathbf x}$ we expand $H_I$ to order $1/|\Rvec|^{5}$:
\begin{multline}
 k^{-1}H_I = [\ra\cdot\rb-3x_A x_B]\frac{1}{|\Rvec|^3}\\
 +\left[3(\ra\cdot\rb)(x_A-x_B)+\frac 3 2(|\ra|^2 x_B-|\rb|^2 x_A)
   +\frac{15}{2}x_A x_B(x_B-x_A)\right]\frac{1}{|\Rvec|^4}\\
 +\Biggl[\frac 3 2 (\ra\cdot\rb)\Bigl((\ra\cdot\rb)-|\ra|^2-|\rb|^2\Bigr)
   +\frac{3}{4}|\ra|^2|\rb|^2
   +\frac{15}{4}\Bigl(2(\ra\cdot\rb)x_A^2+2(\ra\cdot\rb)x_B^2\\
     -|\ra|^2 x_B^2-|\rb|^2 x_A^2
     +2|\ra|^2 x_A x_B+2|\rb|^2 x_A x_B
     -4(\ra\cdot\rb)x_A x_B\Bigr)\\
   +\frac{35}{4}(3x_A^2 x_B^2-2x_A^3 x_B-2x_B^3 x_A)\Biggr]\frac{1}{|\Rvec|^5} + O(|\Rvec|^{-6}).
   \label{eq:H_I-brute-force}
\end{multline}
The first order correction to the ground state energy is the expectation value
of this expanded operator in the unperturbed ground state $|0\rangle$ of Eq.~(\ref{productform}).
By inversion symmetry, the expectation value of all the $1/|\Rvec|^3$ and $1/|\Rvec|^4$ terms
vanish. The non-zero expectation values are (note that these hold in general, not just for the
Drude model)
\begin{align}
 \langle 0|(\ra\cdot\rb)^2|0\rangle &= d a^4,
 & \langle 0|(\ra\cdot\rb)x_A x_B|0\rangle &= a^4,\\
 \langle 0||\ra|^2|\rb|^2|0\rangle &= d^2 a^4, 
 & \langle 0|x_A^2 x_B^2|0\rangle  &= a^4,
\end{align}
and
\be
  \langle 0||\ra|^2 x_B^2|0\rangle = \langle 0||\rb|^2 x_A^2|0\rangle = d a^4.
\ee
In deriving these it is useful to note that e.g.
\be
  \langle 0|x_A y_A |0\rangle = 0
\ee
by rotational symmetry. Combining the previous equations we obtain
\be
  \Delta_1 E_0 = \langle 0|H_I|0\rangle = \frac{3(3-d)(5-d)k}{4}\frac{a^4}{|\Rvec|^5}+O(|R|^{-6}),
\ee
in agreement with \eqref{Delta1E0}. We observe that the first order 
correction is non vanishing in one and two dimensions. 

\section{The $1/|\Rvec|^7$ dependence of $\Delta_1E_0$}
\label{app:first-order-seven}

We first calculate the $1/|\rvec|^5$ term of $V_A(\rvec)$:
\be
  V_A(\rvec) = -\frac{(d-3)}{2}\left(\frac{a^2}{|\rvec|^3}
   +\frac{(d-5)\alpha a^4}{4|\rvec|^5}\right) + O(|\rvec|^{-7}),
  \label{eq:VA-subleading}
\ee
where the positive coefficient $\alpha$ depends on the shape of the wave function and is 
defined by
\be
  \langle 0_A|x_A^4| 0_A\rangle = \alpha a^4.
\ee
In the Drude model we have $\alpha = 3$.
To get the expression \eqref{eq:VA-subleading} for $d>1$ one needs the relation
\be
  \langle 0_A|x_A^4| 0_A\rangle = 3\langle 0_A|x_A^2 y_A^2| 0_A\rangle
\ee
which follows by doing the spherical integration, or by expanding the identity
\be
  \langle 0_A|x_A^4| 0_A\rangle 
  = \left\langle 0_A\middle| \left(\frac{x_A+y_A}{\sqrt 2}\right)^4 \middle| 0_A\right\rangle.
\ee
Plugging \eqref{eq:VA-subleading} into \eqref{eq:D1-VA} we obtain
\be
  \Delta_1E_0 = \frac{(3-d)(5-d)k}{4}\left(\frac{3a^4}{|\Rvec|^5}
    +\frac{5(7-d)\alpha a^6}{2|\Rvec|^7}\right) + O(|\Rvec|^{-9}).
\ee

\end{document}